\documentclass[twocolumn,preprintnumbers,floatfix,aps,prd,amssymb,floatfix,nofootinbib,balancelastpage,superscriptaddress,amsmath]{revtex4-1}

\usepackage[utf8]{inputenc}
\usepackage{amssymb}
\usepackage{amsmath}
\usepackage{amsfonts}
\usepackage{graphicx}
\usepackage{color}
\usepackage{xspace}
\usepackage{comment}
\usepackage{hyperref}
\usepackage[normalem]{ulem}
\usepackage[section]{placeins}
\usepackage{afterpage}
\usepackage{slashed}
\usepackage{enumerate}
\usepackage{enumitem}
\usepackage{caption}
\usepackage{subfigure}
\usepackage{float}
\usepackage{ulem}
\usepackage{verbatim}
\usepackage{multirow,rotating}
\usepackage[dvipsnames]{xcolor}
\usepackage{braket}
\usepackage{silence}
\definecolor{dcolour}{rgb}{.5, .5, .5}
\def\gsim{\raise0.3ex\hbox{$\;>$\kern-0.75em\raise-1.1ex\hbox{$\sim\;$}}}
\def\lsim{\raise0.3ex\hbox{$\;<$\kern-0.75em\raise-1.1ex\hbox{$\sim\;$}}}
\def\gsim{\raise0.3ex\hbox{$\;>$\kern-0.75em\raise-1.1ex\hbox{$\sim\;$}}}
\def\lsim{\raise0.3ex\hbox{$\;<$\kern-0.75em\raise-1.1ex\hbox{$\sim\;$}}}

\newcommand{\ba}[1]{\begin{eqnarray} \label{(#1)}}
\newcommand{\ea}{\end{eqnarray}}

\begin{document}

\title{The strong decay of \texorpdfstring{$Y(4230)\to J/\psi f_0(980)$}{Lg} in light cone sum rules}
\author{\textsc{Yiling Xie}}
\email{xieyl_9@mail.dlut.edu.cn}
\author{\textsc{Hao Sun}}
\email{haosun@dlut.edu.cn}
\affiliation{Institute of Theoretical Physics, School of Physics, Dalian University of Technology, \\ 
No.2 Linggong Road, Dalian, Liaoning, 116024, People’s Republic of China}

\begin{abstract}
In this work, we assign the tetraquark state for $Y(4230)$ resonance,
and investigate the mass and decay constant of $Y(4230)$ in the framework of SVZ sum rules through a different calculation technique. 
Then we calculate the strong coupling $g_{Y J/\psi f_0}$ by considering soft-meson approximation techniques, within the framework of light cone sum rules. 
And we use strong coupling $g_{Y J/\psi f_0}$ to obtain the width of the decay $Y(4230)\to J/\psi f_0(980)$.
Our prediction for the mass is in agreement with the experimental measurement,
and that for the decay width of $Y(4230)\to J/\psi f_0(980)$ is within the upper limit.
\vspace{0.5cm}
\end{abstract}
\maketitle
\setcounter{footnote}{0}

\section{Introduction}
\label{I}

$Y(4230)$ a.k.a. $\psi(4230)$ is the first observed $Y$ state
which was detected through initial-state-radiation (ISR) technique
in the process $e^+e^-\to$ $J/\psi \pi^+\pi^-$
by the BABAR experiment in 2005 \cite{BaBar:2005hhc},
and then confirmed by the CLEO \cite{CLEO:2006tct}
and the Belle \cite{Belle:2007dxy} in the same process.
An accumulation of events with similar characteristics was reported in other two processes
$e^+e^-\to$ $J/\psi\pi^0\pi^0$ (and $J/\psi K^+K^-$)
by the CLEO \cite{CLEO:2006ike}, and also in the decay of
$B^-\to$ $\psi$(4230)$K^-\to J/\psi \pi^+\pi^- K^-$ by BABAR \cite{BaBar:2012vyb} collaboration.

In 2017, the BESIII collaboration has announced a new precise measurement of 
the $e^+e^-\to J/\psi \pi^+\pi^-$ cross section \cite{BESIII:2016bnd},
reporting updated values for the mass and the width of the $Y(4230)$.
Particularly, a second resonance is also presented in the $J/\psi\pi^+\pi^-$ mass spectrum. 
The values of the two observed resonances are, respectively: 
$(4222.0\pm3.1\pm1.4)$ and $(44.1\pm4.3\pm2.0)$ MeV for $Y(4230)$;
$(4320.0\pm10.4\pm7)$ and $(101.4^{+25.3}_{-19.7}\pm10.2)$ MeV for, namely, $Y(4360)$. 
Although Ref.\cite{BESIII:2016bnd} proposes that the structure around 4260 MeV
could be read as a superposition of these two resonances,
and Ref.\cite{BESIII:2018iea} further suggests them as $Y(4230)$ and $Y(4360)$ respectively,
this discussion has not yet settled.
Here in our paper we still concentrate on the $Y(4230)$ resonsant state instead of discussing the combined structure.

Experimentally the $Y(4230)$ is directly produced in $e^+e^-$ annihilation, 
its spin-parity quantum number should be $J^{PC}=1^{--}$, 
which is consistent with that of a vector charmonium state. 
Theorists have tried to categorize it into the vector charmonium group.
However, on account of its mass does not fit 
any mass of charmonium states in the same mass region, 
and that $c\bar{c}$ mainly decay to $D^{(*)}\bar{D}^{(*)}$, 
but the observed $Y$ in such decay does not match the peaks
in the $e^+e^-\to D^{(*)\pm}D^{(*)\mp}$ cross sections 
measured by the BABAR \cite{BESIII:2018iea, Belle:2007qxm} 
and the Belle \cite{Belle:2006hvs} collaborations,
the $Y$(4230) does not look like a normal $c\bar{c}$ state.
Furthermore, for the $1^{--}$ radially excited charmoniums, four of the $S$-wave states:
$\psi(1S)$, $\psi(2S)$, $\psi(3S)$ and $\psi(4S)$,
have already been assigned to $J/\psi$, $\psi(3686)$, 
$\psi(4040)$ and $\psi(4415)$ mesons, respectively,
and two of the $D$-wave states $\psi(1D)$ and $\psi(2D)$ 
have been assigned to $\psi(3770)$ and $\psi(4160)$ mesons. 
In addition, the masses of $\psi(5S)$ and $\psi(3D)$ states in the quark model 
are 4.76 and 4.52 GeV, and thus higher than that of the $Y(4230)$.
According to the above analysis, one can reach that $Y(4230)$ may not consistent with any of 
the $1^{--}$ $c\bar{c}$ states \cite{Zhu:2007wz, Klempt:2007cp, Nielsen:2009uh}.

To further explain the structure of $Y(4230)$,
many theoretical interpretations have been to emerge,
including a tetraquark state \cite{Maiani:2005pe,Zhu:2005hp,Albuquerque:2008up,Ebert:2008kb},
a compact tetraquark state \cite{Maiani:2014aja},
a hadrocharmonium state \cite{Li:2013ssa,Dubynskiy:2008mq}, 
hadronic molecule of $D_0\bar{D}^*$, $D_1 D$, $D_0 D^*$ or $\bar{D} D_1(2420)$ 
\cite{Albuquerque:2008up,Ding:2008gr,Wang:2013cya,Cleven:2013mka},
$\chi_{c1}\omega$ \cite{Yuan:2005dr},
$\chi_{c1}\rho$ \cite{Liu:2005ay},
$J/\psi K\bar{K}$ \cite{MartinezTorres:2009xb},
$\psi^\prime f_0(980)$\cite{Guo:2008zg},
a $c\bar{c}$-gluon hybrid \cite{Zhu:2005hp,Close:2005iz,Kou:2005gt}, 
a charm baryonium \cite{Qiao:2005av}, 
a coupled-channel model \cite{vanBeveren:2006ih,vanBeveren:2009jk}, etc.
However, within the available experimental data, 
none of these theoretical interpretations can be completely 
accepted or excluded to the nature of $Y(4230)$.

For example, if it's in the compact tetraquark model \cite{Maiani:2013nmn},
there exists an isospin-violating process $Y(4230)\to\eta\pi^0J/\psi$
with a sizeable decay width, where both $\pi^0J/\psi$
and $\eta J/\psi$ can be produced from $Z_c^0$ decay.
Therefore, the interpretation of $Y(4230)$ in the compact tetraquark model can
lead to a peak in $e^+e^-\to\eta\pi^0J/\psi$ cross section
and a very prominent peak should appear in $\eta J/\psi$ mass spectrum
between the $D\bar{D}^*$ thereshold \cite{Wu:2013onz}.
However, using the datas from the BESIII experiment 
searching for isospin-violating \cite{BESIII:2015aym},
no $Y(4230)$ signal is observed.
Besides, whatever it is in the compact tetraquark model
it should have isospin and SU(3)-multiplet partner states.
But none of those partners for $\psi$(4230) has been observed in experiments so far.
If $\psi$(4230) is a hadrocharmonium,
it's structure would be formed with the mixing of another hadrocharmonia.
These two hadrocharmonia states contain
spin 1 and spin 0 compact $c\bar{c}$ cores respectively \cite{Li:2013ssa}.
However, based on BESIII data \cite{BESIII:2016bnd}
the decay rate of $Y(4230)$ to non-$J/\psi(h_c)$ charmonium states
should be suppressed \cite{Li:2013ssa}, 
indicating that the above suggestion may not consistent.
If we assign the $\bar{D}D_1(2420)$ molecule to $\psi$(4230),
the binding energy being about 66 MeV is rather large,
though this possibility is not excluded \cite{Brambilla:2019esw}.
There are other candidates for $\psi$(4230),
e.g., $\bar{D}D_1(2430)$, $\bar{D}^*D_0$, $\bar{D}_{s0}D_s^*$, $\bar{D}_{s}D_{s1}$,
whose open charm thresholds are around 4.26 GeV with $J^{PC}=1^{--}$.
Unfortunately, beside $\bar{D}_{s0}D_s^*$, their widths are too broad to make a bound state
so that could not be consistent with the total decay width of
$Y(4230)$ \cite{Wang:2013cya,ParticleDataGroup:2012pjm}.
For the $\bar{D}_{s0}D_s^*$ molecular, its mass is $4.42\pm0.10$ GeV,
which should also be excluded \cite{Albuquerque:2008up}.
Anyway, $Y$(4230) seems not a hadronic molecule.

The $Y(4230)$ may also be assumed as a charmonium hybrid meson.
However, in Ref.\cite{Guo:2008yz} the authors find that the mass of $1^{--}$ hybrid states 
lie at $4.47$ GeV, heavier than the mass of the $Y(4230)$. 
In non-relativistic EFTs, the mass of $Y(4230)$ may consistent with one state of $H_1$ hybrid multiplet,
but $Y(4230)$ disfavors the hybrid interpretation
since it decays to spin triplet charmonium while $H_1$ is only a spin singlet \cite{Berwein:2015vca}.
In Ref.\cite{Ma:2019hsm}, it was found that the color halo picture is compatible with $Y(4260)$ decay properties,
and suggests LHCb and BelleII to search for $(0,1,2)^{+-}$ charmonium-like hybrids in $\xi_{c0,1,2}\eta$ and $J/\psi\omega(\psi)$ final states.
We should not jump to the conclusion that $Y(4230)$ can not be the hybrid state.

In summary, at this point the structure of $Y(4230)$ is not yet fully settled.

In this paper we investigate the strong decay of $Y(4230)\to$ $J/\psi f_0(980)$
observed in the process $e^+e^-\to J/\psi \pi^+\pi^-$.
Notice $Y(4230)$ will decay into a $f_0(980)$
which fits the $S$-wave $[sq][\bar{s}\bar{q}]$ hypothesis \cite{Maiani:2004uc}.
Furthermore, being a member of vector charmonium family
suggests $\psi$(4230) a $[cs][\bar{c}\bar{s}]$ composition.
We therefore consider $Y(4230)$ as a tetraquark state, as in Ref.\cite{Dubnicka:2020xoh}.
This differs from other ideas, i.e., in Ref.\cite{Albuquerque:2011ix}
the $Y(4230)$ was suggested to be a $J/\psi f_0(980)$ bound system. 
We calculate the strong coupling $g_{Y J/\psi f_0}$
by using the method of light cone sum rules, 
with the interpolating current taken from Ref.\cite{Albuquerque:2008up}.
We evaluate the mass of $Y(4230)$ through 
a different calculation technique developed in Ref.\cite{Azizi:2018duk},
not the usual way in two-point sum rules \cite{Agaev:2017tzv}.
Compare the mass prediction of $Y(4230)$ with the result in PDG \cite{ParticleDataGroup:2020ssz},
we confirm our technique generalization is credible.
We then extend it to evaluate the decay constant of $Y(4230)$
which will be used in the numerical calculation of the strong coupling $g_{YJ/\psi f_0}$.
Finally, the decay width of $Y(4230)\to J/\psi f_0(980)$ is obtained,
and further results are compared with the experimental measurement and discussed.

Our work is organized as follows:

In Section.\ref{II} we calculate the mass and decay constant of the $Y(4230)$ state
within two-point sum rule approach develop by Shifman, Vainshtein 
and Zakharov (SVZ sum rules) \cite{Shifman:1978by}.
And we also calculate the strong coupling $g_{YJ/\psi f_0}$
will be derived with the approach of light cone sum rules.
The numerical results and discussions are shown in Section.\ref{III}.
We reach our summary in Section.\ref{sec:summary}

\section{Calculation Framework}
\label{II}

\subsection{ The mass and the decay constant of \texorpdfstring{$Y(4230)$}{Lg} } 

We begin by calculating the mass and the decay constant using the two-point correlation function:
\begin{equation}\label{corm}
	\begin{aligned}
	\Pi_{\mu\nu}(p)=i\int \mbox{d}^4xe^{ipx}
	\braket{0|T\{J_{\mu}^{Y}(x)J_{\nu}^{Y\dagger}(0)\}|0},
	\end{aligned}
\end{equation}
where the interpolating currents are given in the following expression:
\begin{equation}
	\begin{aligned}
		J_{\mu}^{Y}(x)&=\frac{\varepsilon_{abc}\varepsilon_{dec}}{\sqrt{2}}\{[s^T_a(x)C\gamma_5c_b(x)][\bar{s}_d(x)\gamma_\mu\gamma_5C\bar{c}_e^T(x)]\\
		&-[s_a^T(x)C\gamma_\mu\gamma_5c_b(x)][\bar{s}_d(x)\gamma_5C\bar{c}_e^T(x)]\},\\
		J^{Y\dagger}_\nu(0)&=\frac{\varepsilon_{abc}\varepsilon_{dec}}{\sqrt{2}}\{-[\bar{c}_b(0)\gamma_5C\bar{s}_a^T(0)][c_e^T(0)C\gamma_\nu\gamma_5 s_d(0)]\\
		&+[\bar{c}_b(0)\gamma_\nu\gamma_5 C\bar{s}_a^T(0)][c_e^T(0)C\gamma_5 s_d(0)]\}.
	\end{aligned}
\end{equation}
As a first step, we calculate the correlation function by inserting a complete set of hadronic states into Eq.\eqref{corm}:
\begin{equation}
	\begin{aligned}
		\Pi^{\text{h}}_{\mu\nu}(p)&=
		\frac{\braket{0|J_\mu^{Y}|Y(p)}
		\braket{Y(p)|J_\mu^{Y\dagger}|0}}{m_Y^2-p^2}
		+\int_{s^\prime}^\infty d \hat{s}\frac{\rho_{\mu\nu}^{\text{h}}(\hat{s})}{\hat{s}-p^2},
	\end{aligned}
\end{equation}
where the higher resonances and continuous states are represented by $\rho_{\mu\nu}^{\text{h}}(\hat{s})$.
Due to the fact that they would disappear following the Borel transformation, the subtraction terms are not displayed.
We define the decay constant $f_Y$ according to
\begin{equation}
	\begin{aligned}
		\braket{0|J_\mu^{Y}|Y(p)}=m_Yf_Y\epsilon_\mu
	\end{aligned}
\end{equation}
with $\epsilon_\mu$ being the polarization vector of $Y(4230)$. 
After performing the polarization sum equation we can obtain
\begin{equation}\label{POLE-CONT}
	\begin{aligned}
	\Pi_{\mu\nu}^{\text{h}}(p)=\frac{m_Y^2f_Y^2}{m_Y^2-p^2}(-g_{\mu\nu}
	+\frac{p_\mu p_\nu}{m_Y^2})
	+\int_{s^\prime}^\infty d\hat{s}\frac{\rho_{\mu\nu}^{\text{h}}(\hat{s})}{\hat{s}-p^2}.
	\end{aligned}
\end{equation}
On the right side of Eq.\eqref{POLE-CONT}, we are beginning to observe a pole. 
The Borel transformation can be performed on Eq.\eqref{POLE-CONT} to remove the pole, which yields
\begin{equation}
	\begin{aligned}
		&\Pi_{\mu\nu}^{\text{h}}(M^2)=\\
		&m_Y^2f_Y^2e^{-m_Y^2/M^2}(-g_{\mu\nu}+\frac{p_\mu p_\nu}{m_Y^2})+\int_{s^\prime}^\infty d\hat{s}\rho_{\mu\nu}^{\text{h}}(\hat{s})e^{-\hat{s}/M^2}.
	\end{aligned}
\end{equation}

Next, let's consider the correlation function in the OPE side. 
Following Wick Theorm contraction of the heavy and light quarks, we obtain:
\begin{equation}\label{propagator}
	\begin{aligned}
		&\Pi_{\mu\nu}^{\text{OPE}}(p)=\frac{1}{2}\int \mbox{d}^4xe^{ipx}\epsilon_{abc}\epsilon_{dec}\epsilon_{a^\prime b^\prime c^\prime}\epsilon_{d^\prime e^\prime c^\prime}\\
		&-\text{Tr}[\tilde{S}^{aa^\prime}_s(x)\gamma_5S_c^{bb^\prime}(x)\gamma_5]\text{Tr}[S^{dd^\prime}_s(-x)\gamma_\mu\gamma_5\tilde{S}_c^{ee^\prime}(-x)\gamma_\nu\gamma_5]\\
		&+\text{Tr}[\tilde{S}^{aa^\prime}_s(x)\gamma_5S_c^{bb^\prime}(x)\gamma_\nu\gamma_5]\text{Tr}[S^{dd^\prime}_s(-x)\gamma_\mu\gamma_5\tilde{S}_c^{ee^\prime}(-x)\gamma_5]\\
		&+\text{Tr}[\tilde{S}^{aa^\prime}_s(x)\gamma_\mu\gamma_5S_c^{bb^\prime}(x)\gamma_5]\text{Tr}[S^{dd^\prime}_s(-x)\gamma_5\tilde{S}_c^{ee^\prime}(-x)\gamma_\nu\gamma_5]\\
		&-\text{Tr}[\tilde{S}^{aa^\prime}_s(x)\gamma_\mu\gamma_5S_c^{bb^\prime}(x)\gamma_\nu\gamma_5]\text{Tr}[S^{dd^\prime}_s(-x)\gamma_5\tilde{S}_c^{ee^\prime}(-x)\gamma_5],
	\end{aligned}
\end{equation}
where $\tilde{S}_q^{ab}(x)$ represent $CS_q^{ab}(x)C$.
We accept the following expression for propagators of the $u$, $d$, and $s$ quarks in coordinate-space\cite{Huang:2010dc,Agaev:2016mjb}
\begin{equation}
	\begin{aligned}
		S_{q,ab}(x)&=\frac{i\delta_{ab}\slashed{x}}{2\pi^2x^4}-\frac{\delta_{ab}m_q}{4\pi^2x^2}-\frac{{\langle \bar{q}q\rangle }}{12}\\
                &-\frac{i}{32\pi^2}\frac{\lambda^n}{2}g_sG_{\mu\nu}^n\frac{1}{x^2}(\sigma^{\mu\nu}\slashed{x}+\slashed{x}\sigma^{\mu\nu})\\
                &+\frac{i\delta_{ab}\slashed{x}m_q{\langle \bar{q}q\rangle }}{48}
                -\frac{\delta_{ab}{\langle \bar{q}g_s\sigma Gq \rangle}x^2}{192}\\
                &+\frac{i\delta_{ab}x^2\slashed{x}m_q\langle\bar{q}g_s\sigma Gq\rangle}{1152}\\
                &-\frac{i\delta_{ab}x^2\slashed{x}g_s^2\langle\bar{q}q\rangle^2}{7776}-\frac{\delta_{ab}x^4\langle\bar{q}q\rangle\langle g_s^2GG\rangle}{27648}.
	\end{aligned}
\end{equation}
The heavy quark propagator is given, in terms of the second kind Bessel functions $K_v(x)$, as \cite{Aliev:2016xvq}:
\begin{equation}\label{Bpropagator}
	\begin{aligned}
		&S^{ab}_{c}(x)=\frac{m_Q^2}{4\pi^2}[\frac{K_1(m_Q\sqrt{-x^2})}{\sqrt{-x^2}}\delta_{ab}+i\frac{\slashed{x}K_2(m_Q\sqrt{-x^2})}{(\sqrt{-x^2})^2}\delta_{ab}]\\
&-\frac{g_sm_Q}{16\pi^2}\int_0^1\mbox{d}vG^ {\mu\nu}_{ab}(vx)[(\sigma_{\mu\nu}\slashed{x}+\slashed{x}\sigma_{\mu\nu})\frac{K_1(m_Q\sqrt{-x^2})}{\sqrt{-x^2}}\\
&+2\sigma^{\mu\nu}K_0(m_Q\sqrt{-x^2})].
	\end{aligned}
\end{equation}
Notice the heavy quark propagator here is different from the expression presented in the usual way, 
for example in Ref.\cite{Agaev:2017tzv}, where the heavy quark propagator is expressed in the momentum space.
If use momentum expression of propagator in Eq.\eqref{propagator}, we have to face divergences in the double integrals like
\begin{equation}
	\begin{aligned}
		\int\frac{d^4k_1d^4k_2}{(k_1^2-m_c^2)(k_2^2-m_c^2)}.
	\end{aligned}
\end{equation}
As shown in Ref.\cite{Azizi:2018duk}, by using an appropriate representation of 
the modified Bessel functions in heavy quark propagator, like in Eq.\eqref{Bpropagator},{}
results without any divergences can be obtained. 
Since here we are using the SVZ sum rules instead of the LCSR,
we have to modify the calculation when the paticle distribution function do not participate in Eq\eqref{propagator}.
We showe the details of the modification in Appendix \ref{C}.

The correlation function $\Pi^{\text{OPE}}_{\mu\nu}(p)$ has also the following decomposition over the Lorentz structures
\begin{equation}
	\begin{aligned}
		\Pi^{\text{OPE}}_{\mu\nu}(p)=\Pi^{\text{OPE}}(p)g_{\mu\nu}+\widetilde{\Pi}^{\text{OPE}}(p)p_\mu p_\nu,
	\end{aligned}
\end{equation}
and we choose to work with the term $\sim g_{\mu\nu}$,
which can be represented as the dispersion integral
 \begin{equation}
 	\begin{aligned}
 		\widetilde{\Pi}^{\text{OPE}}(p)=\int_{4m_c^2}^\infty\mbox{d}\hat{s}\frac{\rho^{\text{OPE}}(\hat{s})}{\hat{s}-p^2},
 	\end{aligned}
 \end{equation}
where $\rho^{\text{OPE}}(\hat{s})$ is the corresponding spectral density.

The Borel transformation and the quark-hadron duality can be applied to $\widetilde{\Pi}^{\text{OPE}}(p)$ to obtain:
\begin{equation}
	\begin{aligned}
		&p_\mu p_\nu\int_{4m_c^2}^\infty\mbox{d}\hat{s}\rho^{\text{OPE}}(\hat{s})e^{-\hat{s}/M^2}=\\
		&m_Y^2f_Y^2e^{-m_Y^2/M^2}(\frac{p_\mu p_\nu}{m_Y^2})+p_\mu p_\nu\int_{s_0}^\infty\rho^{\text{h}}(\hat{s})e^{-\hat{s}/M^2}.
 	\end{aligned}
\end{equation}
Next, take out the contribution from the continuum to get:
\begin{equation}
	\begin{aligned}
		f_Y^2e^{-m_Y^2/M^2}=\int_{4m_c^2}^{s_0}\mbox{d}\hat{s}\rho^{\text{OPE}}(\hat{s})e^{-\hat{s}/M^2}.
 	\end{aligned}
\end{equation}
The $Y(4230)$ state mass can be determined by the sum rule
\begin{equation}
	\begin{aligned}
		m_Y^2=\frac{\int_{4m_c^2}^{s_0}\mbox{d}\hat{s}\hat{s}\rho^{\text{OPE}}(\hat{s})e^{-\hat{s}/M^2}}{\int_{4m_c^2}^{s_0}\mbox{d}\hat{s}\rho^{\text{OPE}}(\hat{s})e^{-\hat{s}/M^2}}.
	\end{aligned}
\end{equation}

\subsection{The strong coupling \texorpdfstring{$g_{YJ/\psi f_0}$}{Lg} in light cone sum rules}

It is necessary to calculate the strong coupling $g_{YJ/\psi f_0}$ first, based on the light cone sum rules(LCSR),
before predicting the width of $Y(4230)\to J/\psi f_0(980)$.
We begin by using the two-point correlation function:
\begin{equation}\label{3}
	\begin{aligned}
	\Pi_{\mu\nu}(p^\prime,q)=i\int \mbox{d}^4xe^{ipx}
	\braket{f_0(q)|T\{J_{\mu}^{J/\psi}(x)J_{\nu}^{Y\dagger}(0)\}|0},
	\end{aligned}
\end{equation}
where $f_0$ represents the scalar meson $f_0(980)$. $Y(4230)$ has momentum $p^\prime=p+q$,
and $p$, $q$ represent the four-momentum for $J/psi$ and $f_0$, respectively.
$J_{\mu}^{J/\psi}$ is the interpolating current of $J/\psi$ given by \cite{Albuquerque:2008up, Becirevic:2013bsa}
\begin{equation}
	\begin{aligned}
	&J_{\mu}^{J/\psi}(x)=\bar{c}_i(x)\gamma_\mu c_i(x).
	\end{aligned}
\end{equation}
here $i$ denotes the color indexes and $C$ is the charge conjugation matrix.

\subsubsection{Phenomenological side calculation}
Next, we must build a relationship between the correlation function $\Pi_{\mu\nu}(p^\prime,q)$ and the strong coupling $g_{YJ/\psi f_0}$.

By adding two complete sets of hadronic states to Eq.\eqref{3},
we are able to construct the phenomenological expression of the correlation function:
\begin{equation}\label{pheno}
	\begin{aligned}
		&\Pi^{\text{h}}_{\mu\nu}(p^\prime,q)=\\
		&\frac{\braket{0|J_{\mu}^{J/\psi}|J/\psi(p)}
		       \braket{f_0(q)J/\psi(p)|Y(p^\prime)}
			 \braket{Y(p^\prime)|J_{\nu}^{Y\dagger}|0}}
			 {(p^{\prime2}-m_{Y}^2)(p^2-m_{J/\psi}^2)}\\
			 &+\int_{s_1^\prime}^\infty\int_{s_2^\prime}^\infty
			   \frac{ds_1ds_2\ \rho_{\mu\nu}^{\text{h}}(s_1,s_2)}{(s_1-p^2)(s_2-p^{\prime2})}
			   +\cdots,
	\end{aligned}
\end{equation}
where $\rho_{\mu\nu}^{\text{h}}(s_1,s_2)$ represents the contributions of the continuum states and higher resonances,
the lowest continuum state thresholds are indicated by the symbols $s_1^\prime$ and $s_2^\prime$.

By parameterizing the hadronic matrix element 
\begin{equation}\label{decay constants}
	\begin{aligned}
		&\braket{0|J_{\mu}^{J/\psi}|J/\psi(p)}=m_{J/\psi}f_{J/\psi}\varepsilon_\mu, \\
		&\braket{Y(p^\prime)|J_{\nu}^{Y\dagger}|0}=m_{Y}f_{Y}\varepsilon_\nu^{\prime*},\\
		&\braket{f_0(q)J/\psi(p)|Y(p^\prime)}=g_{YJ/\psi f_0}\\
		\times &((p^\prime\cdot p)(\varepsilon^*\cdot \varepsilon^\prime)-(p^\prime\cdot \varepsilon^*)(p\cdot\varepsilon^\prime)),
	\end{aligned}
\end{equation}
and perform the polarization sum, we can easily show that
\begin{equation}
	\begin{aligned}\label{1}
		\Pi^{\text{h}}_{\mu\nu}(p^\prime,q)&=\frac{m_{J/\psi}m_{Y}f_{J/\psi}f_{Y}g_{YJ/\psi f_0}}{(p^2-m_{J/\psi}^2)(p^{\prime2}-m_{Y}^2)}\\
		&\times [ \frac{(m_{J/\psi}^2+m_Y^2)}{2}g_{\mu\nu}-p^\prime_\mu p_\nu ] +\cdots\\
		&=\Pi^{\text{Phys}}(p^\prime,q)g_{\mu\nu}+\widetilde{\Pi}^{\text{Phys}}(p^\prime,q)p^\prime_\mu p_\nu,
	\end{aligned}
\end{equation}
where $m_{J/\psi}$ and  $m_{Y}$ are the mass of $J/\psi$ and $Y(4230)$ respectively.
$\varepsilon$ and $\varepsilon^\prime$ denote the polarization vector of
the $J/\psi$ and $Y(4230)$, respectively.
$g_{YJ/\psi f_0}$ is the invariant constant parameterizing the hadronic matrix element.

In this study, we choose to proceed with a structure that is proportional to $g_{\mu\nu}$
\begin{equation}\label{CF}
	\begin{aligned}
		\Pi^{\text{h}}(p^\prime,q)&=\frac{m_{J/\psi}m_{Y}f_{J/\psi}f_{Y}g_{YJ/\psi f_0}m^2}{(p^2-m_{J/\psi}^2)(p^{\prime2}-m_{Y}^2)}\\
		&+\int_{s_1^\prime}^\infty\int_{s_2^\prime}^\infty\frac{\mbox{d}s_1\mbox{d}s_2\rho^{\text{h}}(s_1,s_2)}{(s_1-p^2)(s_2-p^{\prime2})}+\cdots,
	\end{aligned}
\end{equation}
where we define $m^2$ $=\frac{m_{J/\psi}^2+m_Y^2}{2}$.
The correlation function Eq.\eqref{CF} can be transformed into an equation below
by applying the Borel transformations to the variables $p^2$ and $p^{\prime2}=(p+q)^2$,
\begin{equation}\label{ha}
	\begin{aligned}
		&\mathcal{B}_{p^2}(M_1^2)\mathcal{B}_{p^{\prime 2}}(M_2^2)\Pi^{\text{h}}(p^\prime,q)=\\
		&m_{J/\psi}m_{Y}f_{J/\psi}f_{Y}g_{YJ/\psi f_0}m^2\exp[-\frac{m_{J/\psi}}{M_1^2}-\frac{m_{Y}}{M_2^2}]\\
		&+\int_{s_1^\prime}^\infty\int_{s_2^\prime}^\infty\mbox{d}s_1\mbox{d}s_2 \exp[-\frac{s_1}{M_1^2}-\frac{s_2}{M_2^2}] \rho^{\text{h}}(s_1,s_2).
	\end{aligned}
\end{equation}

Since we have the following formula for a general dispersion relation:
\begin{equation}\label{dispersion}
	\begin{aligned}
		\Pi_{\mu\nu}(p^\prime,q)=\frac{1}{\pi^2}\int\int \frac{ds_1ds_2\text{Im}\Pi_{\mu\nu}(s_1,s_2)}{(s_1-p^2)(s_2-p^{\prime2})}+\cdots,
	\end{aligned}
\end{equation}
where the subtraction terms and single dispersion integrals aren't provided
due to they'll all vanish when the double Borel transformation is applied to Eq.\eqref{dispersion}.
By choosing to proceed with a structure that is proportional to $g_{\mu\nu}$, we can represent the OPE result for the correlation function as
\begin{equation}
	\begin{aligned}
	\Pi^{\text{OPE}}(p^\prime,q)
	=\int\int \frac{ds_1ds_2\ \rho^{\text{OPE}}(s_1,s_2)}{(s_1-p^2)(s_2-p^{\prime2})},
	+\cdots
	\end{aligned}
\end{equation}
where
\begin{equation}
	\begin{aligned}
	\rho_{\mu\nu}^{\text{OPE}}(s_1,s_2)=\frac{\text{Im}\Pi(s_1,s_2)}{\pi^2}.
	\end{aligned}
\end{equation}
After performing the Borel transformations, we can derive
\begin{equation}\label{ope}
	\begin{aligned}
	\Pi^{\text{OPE}}&(M_1^2,M_2^2)=\\
	&\int\int\mbox{d}s_1\mbox{d}s_2 \exp[-\frac{s_1}{M_1^2}-\frac{s_2}{M_2^2}]
	\rho^{\text{OPE}}(s_1,s_2).
	\end{aligned}
\end{equation}

This is then accomplished by applying the quark-hadron duality,
which allows the integral of the hadronic spectral density to equal that of the OPE spectral density, in a certain region:
\begin{equation}\label{hope}
	\begin{aligned}
		&\int_{s_1^\prime}^\infty\int_{s_2^\prime}^\infty\mbox{d}s_1\mbox{d}s_2 \exp[-\frac{s_1}{M_1^2}-\frac{s_2}{M_2^2}] \rho^{\text{h}}(s_1,s_2)=\\
		&\int_{s_1^0}^\infty\int_{s_2^0}^\infty\mbox{d}s_1\mbox{d}s_2 \exp[-\frac{s_1}{M_1^2}-\frac{s_2}{M_2^2}] \rho^{\text{OPE}}(s_1,s_2) .
	\end{aligned}
\end{equation}
After equating \eqref{ha} and \eqref{ope}, and substituting with \eqref{hope}, we get the following equation for the strong coupling:
\begin{equation}\label{h}
	\begin{aligned}
	&g_{YJ/\psi f_0}=\frac{1}{m_{J/\psi}m_{Y}f_{J/\psi}f_{Y}m^2}\exp[\frac{m_{J/\psi}}{M_1^2}
	+\frac{m_{Y}}{M_2^2}]\\
	&\times\int^{s_1^0}_{4m_c^2}\int^{s_2^{4m_c^2}}_0\mbox{d}s_1\mbox{d}s_2
	\exp[-\frac{s_1}{M_1^2}-\frac{s_2}{M_2^2}] \rho^{\text{OPE}}(s_1,s_2).
	\end{aligned}
\end{equation}

As we can see from Eq.\eqref{3}, since the interpolating currents of $Y(4230)$ and $J/psi$ are located at the point $x$ and $0$, respectively, there will still be a quark element $\braket{f_0(q)|[\bar{s}(0)s(0)]|0}$ after the $\bar{c}$ and $c$ quark fields are contracted.
Because $\braket{f_0(q)|[\bar{s}(x)s(0)]|0}$ will disappears and reduces to normalization factors when $x\to 0$.
This situation can be replace by the kinematical limit $q\rightarrow 0$ which call soft-meson approximation\cite{Belyaev:1994zk}.
Such approximation leads hadronic representation to
\begin{equation}
	\begin{aligned}
		&\Pi^{\text{h}}(p^\prime,q)=\frac{m_{J/\psi}m_{Y}f_{J/\psi}f_{Y}m^2}{(p^2-m^2)^2} g_{YJ/\psi f_0}+\cdots,
	\end{aligned}
\end{equation}
and the Borel transformation on the variable $p^2$ applied to this correlation function yields
\begin{equation}
	\begin{aligned}
		&\Pi^{\text{h}}(p^\prime,q)=m_{J/\psi}m_{Y}f_{J/\psi}f_{Y}\frac{m^2}{M^2}g_{YJ/\psi f_0}e^{-\frac{m^2}{M^2}}+\cdots,
	\end{aligned}
\end{equation}
Following \cite{Ioffe:1983ju,Belyaev:1994zk}, we apply the operator
\begin{equation}
	\begin{aligned}
		(1-\frac{1}{M^2})M^2e^{m^2/M^2}
	\end{aligned}
\end{equation}
on both sides of the sum rules expression to remove unsuppressed contributions,
we can obtain
\begin{equation}\label{couplingn}
	\begin{aligned}
		&g_{YJ/\psi f_0}=\frac{1}{m_{J/\psi}m_{Y}f_{J/\psi}f_{Y}m^2}(1-\frac{1}{M^2})M^2e^{m^2/M^2}\\
		&\times\int^{s_0}_{4m_c^2}\mbox{d}\hat{s}\exp[\frac{m_{J/\psi}}{2M^2}+\frac{m_{Y}}{2M^2}-\frac{\hat{s}}{M^2}] \rho^{\text{OPE}}(\hat{s}),
	\end{aligned}
\end{equation}
which depends only on $\hat{s}$ due to the soft-meson approximation.

\subsubsection{OPE side calculation}
Taking into account that $g_{XJ/\psi \phi}$ has a relationship to the OPE part of the correlation function, we will calculate it.
According to the Wick Theorm, we can derive
\begin{equation}\label{2}
	\begin{aligned}
	\Pi_{\mu\nu}(p^\prime,q)
	&=i\int d^4xe^{ipx}\braket{f_0(q)|T\{J_{\mu}^{J/\psi}(x)J_{\nu}^{Y\dagger}(0)\}|0}\\
	&=i\int d^4xe^{ipx}\frac{\varepsilon_{abc}\varepsilon_{dec}}{\sqrt{2}}\braket{f_0(q)|[\bar{s}^a_\alpha(0) s^d_\beta(0)]|0}\\
	&\times[(\gamma_5)\tilde{S}^{ib}_c(x)\gamma_\mu \tilde{S}^{ie}_c(-x)\gamma_\nu\gamma_5 \\
	&-(\gamma_\nu\gamma_5)\tilde{S}^{ib}_c(x)\gamma_\mu \tilde{S}^{ie}_c(-x)\gamma_5]_{\alpha\beta}.
	\end{aligned}
\end{equation}
For the heavy quark propagator on the light cone we employ
its expression in terms of \cite{Agaev:2017uky}
\begin{equation}\label{Hpropagator}
\begin{aligned}
S^{ab}_{c}(x)=&i\int\frac{d^4k}{(2\pi)^4}e^{-ikx}\\
[&\frac{\delta_{ab}(\slashed{k}+m_c)}{k^2-m^2_c}-\frac{g_sG_{ab}^{\alpha\beta}}{4}\frac{\sigma_{\alpha\beta}(\slashed{k}+m)+(\slashed{k}+m)\sigma_{\alpha\beta}}{(k^2-m^2)^2}],
\end{aligned}
\end{equation}
where we adopt the notation
\begin{equation}
\begin{aligned}
G_{ab}^{\mu\nu}\equiv G_i^{\mu\nu}t^i_{ab},i=1,2,\cdots,8.
\end{aligned}
\end{equation}
Substituting the summation and the expansion
\begin{equation}\label{summation}
\begin{aligned}
\bar{s}^d_\alpha(0) s^{d^\prime}_\beta(0)=\frac{1}{3}\delta_{dd^\prime}\bar{s}_\alpha(0) s_\beta(0),\\
\bar{s}_\alpha(0) s_\beta(0)\equiv\frac{1}{4}\Gamma^j_{\alpha\beta}\bar{s}(0)\Gamma^j s(0),
\end{aligned}
\end{equation}
into Eq.\eqref{2}, we can obtain
\begin{equation}\label{opecorrelation}
\begin{aligned}
\Pi_{\mu\nu}(p^\prime,q)&=i\int \mbox{d}^4xe^{ipx}\frac{\varepsilon_{abc}\varepsilon_{dec}\delta_{ad}}{12\sqrt{2}}\braket{f_0(q)|[\bar{u}(0)\Gamma^j u(0)]|0}\\
	&\times\text{Tr}[(\gamma_5)\tilde{S}^{ib}_c(x)\gamma_\mu \tilde{S}^{ie}_c(-x)\gamma_\nu\gamma_5\\
	&-(\gamma_\nu\gamma_5)\tilde{S}^{ib}_c(x)\gamma_\mu \tilde{S}^{ie}_c(-x)\gamma_5\Gamma^j].
\end{aligned}
\end{equation}
where
\begin{equation}
\begin{aligned}
\Gamma^j=1,\gamma_5,\gamma_\mu,i\gamma_5\gamma_\mu,\frac{\sigma_{\mu\nu}\gamma_5}{\sqrt{2}}.
\end{aligned}
\end{equation}

After substituting the propagator, Using the Particle Distribution Amplitudes (DAs) of $f_0(980)$ in Appendix \ref{appendix:A}
and contracting the color index by the SU(N) algebra
\begin{equation}
	\begin{aligned}
	\varepsilon_{abc}\varepsilon_{dec}\delta_{ad}\delta_{bi}\delta_{ei}=-C_A(1-C_A)=6,
	\end{aligned}
\end{equation}
we will encounter four-dimensional integrals, for example
\begin{equation}\label{four integral}
	\begin{aligned}
		\int\frac{d^4k_1}{(2\pi)^4}\int\frac{d^4k_2}{(2\pi)^4}\frac{e^{-i(k_1-k_2)x}g_{\mu\nu}k_1\cdot k_2}{(k_1^2-m^2_c)(k_2^2-m^2_c)}.
	\end{aligned}
\end{equation}
In Appendix \ref{appendix:B}, we provide main steps to calculate some four-dimensional integrals like \eqref{four integral}.
By choosing the term in proportional to $g_{\mu\nu}$, we can derive
\begin{equation}
	\begin{aligned}
		\rho^{\text{OPE}}(\hat{s})=\frac{m_{f_0}\bar{f}_{f_0}(\hat{s}+2m_c^2)\sqrt{2\hat{s}(\hat{s}-4m_c^2)}}{24\pi^2\hat{s}},
	\end{aligned}
\end{equation}
where $m_{f_0}$ and $\bar{f}_{f_0}$ are the mass and decay constant of $f_0(980)$ respectively. 
The strong coupling is then evaluated by Eq.\eqref{couplingn}.
Besides, we can derive the decay width of $Y(4230)\to J/\psi f_0(980)$ as\cite{Agaev:2016dev}
 \begin{equation}
 	\begin{aligned}
 		&\Gamma(Y(4230)\to J/\psi f_0(980))=\frac{g^2_{YJ/\psi f_0}m^2_{J/\psi}}{24\pi}\\
 		&\times\lambda(m_Y,m_{J/\psi},m_{f_0})\left(3+\frac{2\lambda(m_Y,m_{J/\psi},m_{f_0})}{m^2_{J/\psi}}\right),
 	\end{aligned}
 \end{equation}
where 
 \begin{equation}
 	\begin{aligned}
 		\lambda(a,b,c)=\frac{\sqrt{a^4+b^4+c^4-2*(a^2b^2+b^2c^2+c^2a^2)}}{2a}.
 	\end{aligned}
 \end{equation}

\section{Numerical calculation}
\label{III}

\subsection{Input parameters}

In this section, we present the mass and decay constant of $Y$(4230), 
and analyze the numerical results for the decay width of $Y(4230) \to J/\psi f_0(980)$ as well.
We use the following parameters for the numerical calculation. 
The current charm-quark mass, $m_c=(1.275\pm 0.025)$ GeV, 
the $J/\psi$-meson mass $m_{J/\psi}=(3096.900\pm 0.006)$ MeV 
and $f_0$(980) mass $m_{f_0}=(990\pm 20)$ MeV 
from the Particla Data Group (PDG) \cite{ParticleDataGroup:2020ssz}. 
The $J/\psi$ and $f_0$(980) decay constants are taken 
as $f_{J/\psi}$=$0.405$ GeV \cite{Dias:2013xfa},
$\bar{f}_{f_0}$=$0.18\pm 0.015$ GeV \cite{Colangelo:2010bg}.
The current-quark-mass for the s-quark is $m_s=93^{+11}_{-5}$ MeV from PDG.
In addition, we also need to know the values of the non-perturbative vacuum condensates. 
The related parameters are \cite{Albuquerque:2008up,Narison:2002woh,Narison:1989aq}
\begin{equation}
	\begin{aligned}
		&\langle{\bar{q}q}\rangle=-(0.24\pm 0.01)^3\ \text{GeV}^3,\\
		&\braket{\bar{s}s}=(0.8\pm0.1)\times \braket{\bar{q}q},\\
		&\braket{\frac{\alpha_s}{\pi} GG}=(0.012)\ \text{GeV}^4,\\
		&\braket{g_s\bar{s}\sigma Gs}=m_0^2\times\braket{\bar{s}s},\\
		&m_0^2=0.8\ GeV^2,\\
		&m_c=1.275\pm0.025\ \text{MeV}.\\
	\end{aligned}
\end{equation}

The sum rule predictions depend on two parameters, continuum threshold $s_0$ and borel mass $M^2$.

$s_0$ is being correlated with the first of excited states of $Y(4230)$. However, the experimental results show that there is no resonance activity associated with states of $Y(4230)$. We can naturaly chose $s_0=(m_X+0.5)^2$ $\rm GeV^2$, because the mass gap between the ground state and the first excited state is regularly around $0.5$ GeV in charmonia and bottomonia(Table \ref{table:I}).
\begin{table}[H]
	\centering
	\caption{Quark model masses calculated for the first three levels of charmonia and bottomonia \cite{Olpak:2016wkf}.}
	\resizebox{\linewidth}{!}{
\begin{tabular}{|c|c|c|c|c|c|c|}
\hline 
Masses  &\multicolumn{3}{|c|}{$c\bar{c}$}  & \multicolumn{3}{|c|}{$b\bar{b}$}  \\
\hline \hline 
M(GeV) $\backslash$ n & n=1 & n=2 & n=3 & n=1 & n=2 & n=3 \\
\hline 
$M_{ ^{1} P_{1}}\left(h_{q}\right)$ & 3.53 & 3.96 & 4.37 & 9.88 & 10.3 & 10.6 \\
\hline 
$M_{ ^{3}P_{0}}\left(\chi_{q 0}\right)$ & 3.37 & 3.88 & 4.30 & 9.81 & 10.2 & 10.7 \\
\hline 
$M_{ ^{3} P_{1}}\left(\chi_{q 1}\right)$ & 3.54 & 3.97 & 4.33 & 9.89 & 10.3 & 10.6 \\
\hline 
$M_{ ^{3} P_{2}}\left(\chi_{q 2}\right)$ & 3.54 & 3.98 & 4.34 & 9.89 & 10.3 & 10.6\\
\hline
\end{tabular}
}
\label{table:I}
\end{table}

Additionally, Table \ref{table:II} contains experimental data taken from PDG that support the majority of the computations in Table \ref{table:I}.
\begin{table}[H]
	\centering
	\caption{Masses of experimentally observed states in Particle Data Group listings\cite{ParticleDataGroup:2012pjm}.}
	\resizebox{\linewidth}{!}{
\begin{tabular}{|c|c|c|c|c|c|c|}
\hline Masses & \multicolumn{3}{|c|}{$c \bar{c}$} & \multicolumn{3}{|c|}{$b \bar{b}$} \\
\hline \hline$M(M e V) \backslash n$ & $n=1$ & $n=2$ & $n=3$ & $n=1$ & $n=2$ & $n=3$ \\
\hline$M_{ ^1P_{1}}\left(h_{q}\right)$ & $3525.38$ & $-$ & $-$ & $9899.3$ & $10259.8$ & $-$ \\
\hline$M_{ ^{3} P_{0}}\left(\chi_{q 0}\right)$ & $3414.75$ & $-$ & $-$ & $9859.44$ & $10232.5$ & $-$ \\
\hline$M_{ ^3 P_{1}}\left(\chi_{q 1}\right)$ & $3510.66$ & $-$ & $-$ & $9892.78$ & $10255.46$ & $10512.1$ \\
\hline$M_{ ^3 P_{2}}\left(\chi_{q 2}\right)$ & $3556.20$ & $3922.5$ & $-$ & $9912.21$ & $10268.65$ & $-$ \\
\hline
\end{tabular}
}
\label{table:II}
\end{table}

Moreover, we can refer to the QCD sum rules calculations listed in Table \ref{table:III}. There is a mass difference of $0.4\sim 0.6 $ GeV between 1S and 2S tetraquark states. So we adopt this mass gap and employ 
\begin{equation}
	\begin{aligned}
		(4.23+0.40)^2\ \text{GeV}^2\leq s_0\leq (4.23+0.60)^2\ \text{GeV}^2.
	\end{aligned}
\end{equation}

\begin{table}[H]
	\centering
	\caption{The mass difference between the 1S and 2S hidden-charm tetraquark states with the possible assignments \cite{Wang:2021lkg}.}
	\resizebox{\linewidth}{!}{
	\begin{tabular}{|c|c|c|c|c|}
	\hline \hline $J^{P C}$ & $1 \mathrm{~S}$ & 2 $\mathrm{~S}$ & $\text { Mass difference }$ & $\text { References }$ \\
	\hline $0^{++}$ & $X(3915)$ & $X(4500)$ & $588 \mathrm{MeV}$ & \cite{Lebed:2016yvr,Wang:2016gxp} \\
	\hline $1^{++}$ & $X(4140)$ & $X(4685)$ & 566 $\mathrm{MeV}$ & {\cite{Wang:2021ghk,Wang:2018qpe}} \\
	\hline $1^{+-}$ & $Z_{c}(3900)$ & $Z_{c}(4430)$ & $591 \mathrm{MeV}$ & \cite{Maiani:2014aja,Nielsen:2014mva,Wang:2014vha}\\
	\hline $1^{+-}$ & $Z_{c}(4020)$ & $Z_{c}(4600)$ & $576 \mathrm{MeV}$ & \cite{Chen:2019osl,Wang:2019hnw} \\
	\hline \hline
	\end{tabular}
	}
\label{table:III}
\end{table}

The borel mass $M^2$can be determined by two principles:
\begin{enumerate}
\item
Our requirement is that the high dimension condensates make up not more than $10\%$ of the total contribution to the OPE:
\begin{equation}
	\begin{aligned}
	\mathrm{CVG} \equiv\left|\frac{\Tilde{\Pi}^{\braket{\bar{q}g_sGq}+\cdots}\left( M^{2},\infty\right)}{\Tilde{\Pi}^{OPE}\left(M^{2},\infty \right)}\right| \leq 10\%,
	\end{aligned}
\end{equation}
where ellipsis represent higher dimension contributions.
\item We require that the pole contribution (PC) in Eq.\eqref{POLE-CONT} should exceed $50\%$
\begin{equation}
	\begin{aligned}
		\text{PC}=\frac{\Tilde{\Pi}^{OPE}(M^2,s_0)}{\Tilde{\Pi}^{OPE}(M^2,\infty)}\ge  50\% .
	\end{aligned}
\end{equation}
\end{enumerate}

\begin{figure}[H]
\centering
  \includegraphics[width=7cm]{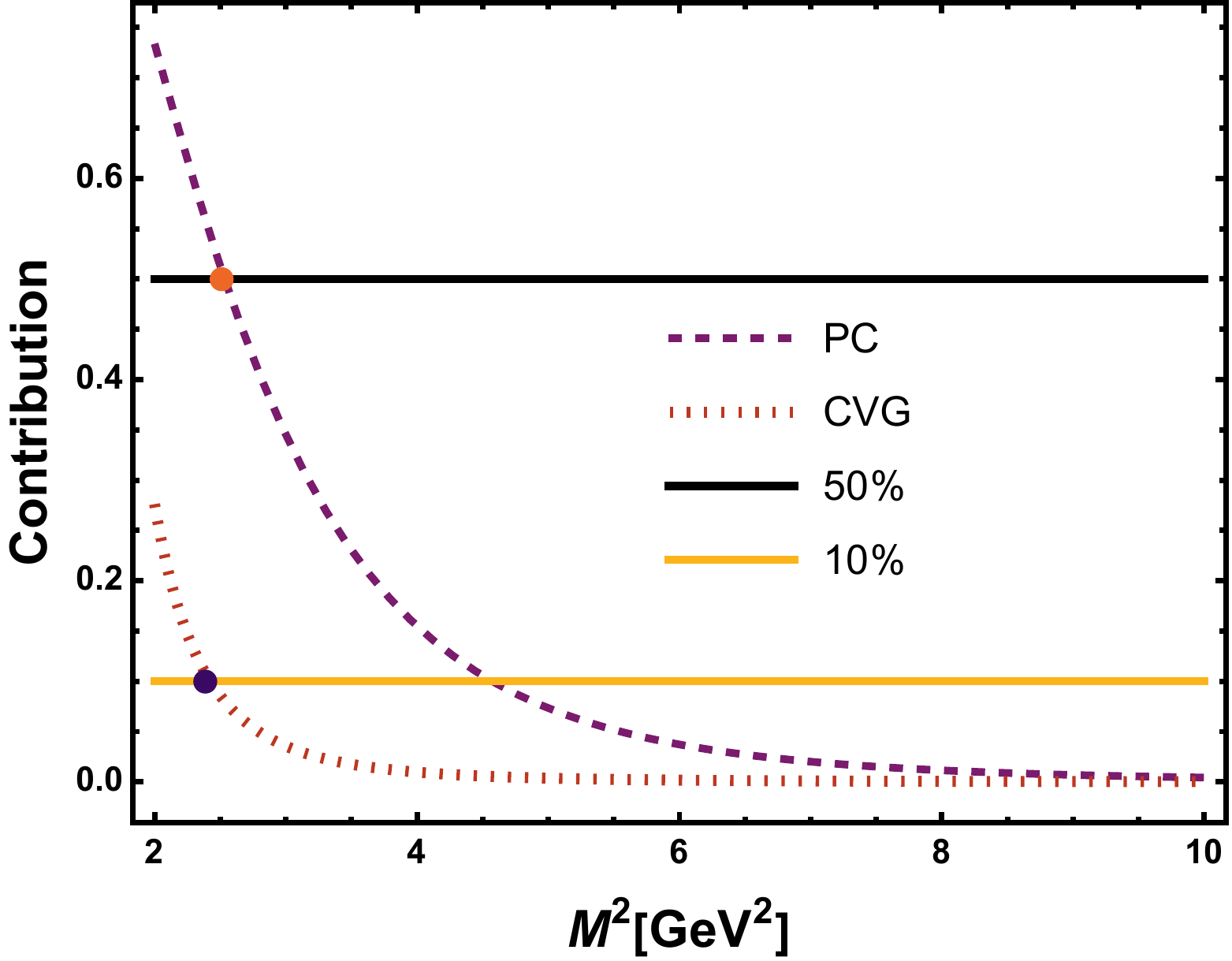}
  \caption{
  Convergence (CVG) and pole contribution (PC) for $\psi(4230)$. }
\label{Fig:fig1}
\end{figure}

As seen in FIG.\ref{Fig:fig1}, the red dot indicates the point at which CVG becomes $50\%$, where the maximum achievable $M^2$ can be attained.
And we can select the minimum $M^2$ from the black dot which present PC converges with $10\%$. Therefore we require the region of the $M^2$ to be
\begin{equation}
	\begin{aligned}
		2.39\ \text{GeV}^2\leq M^2\leq 2.51\ \text{GeV}^2,
	\end{aligned}
\end{equation}

\subsection{The mass, decay constant and decay width}

\begin{figure}[htbp]
	\centering
  \includegraphics[width=7cm]{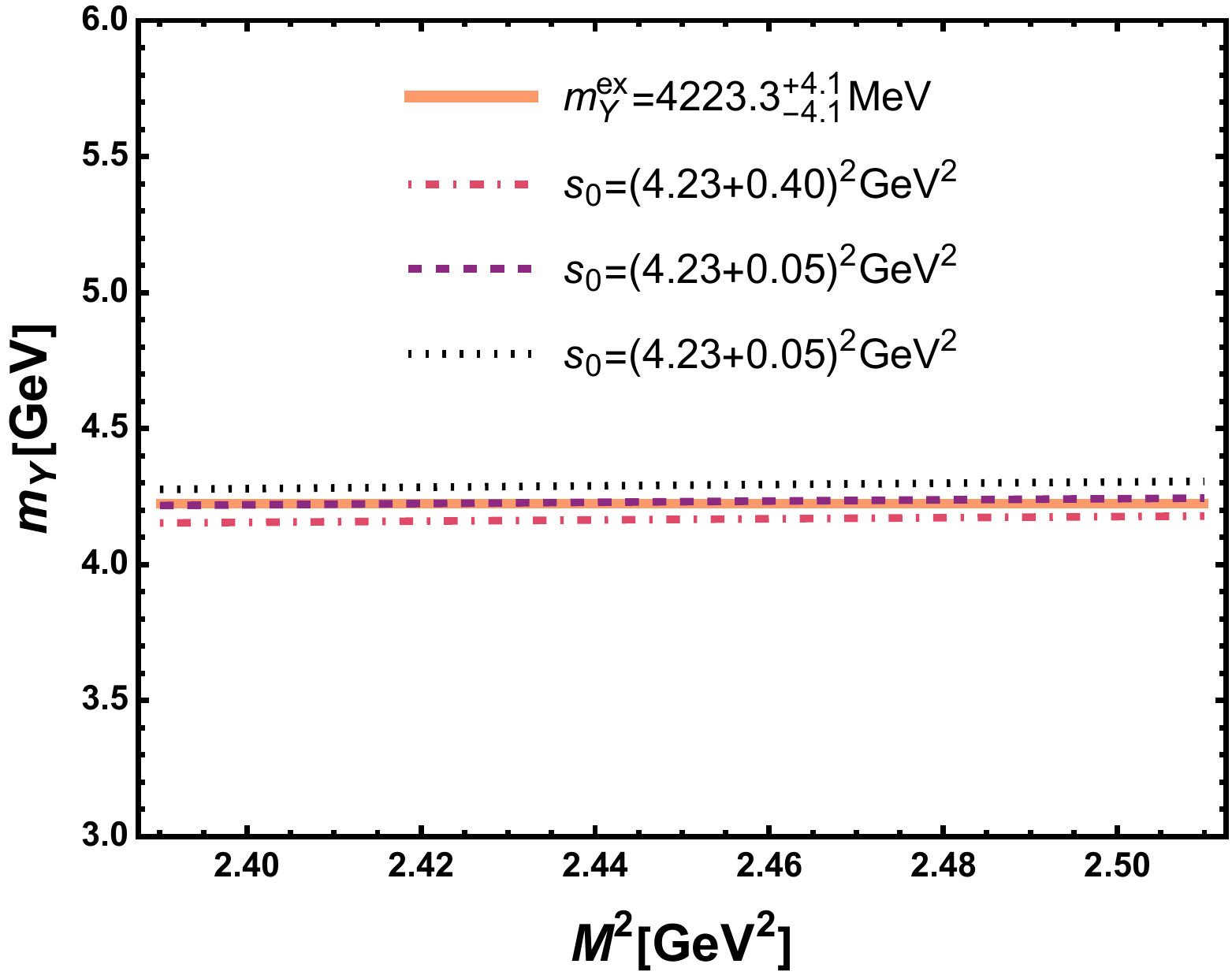}
  \includegraphics[width=7cm]{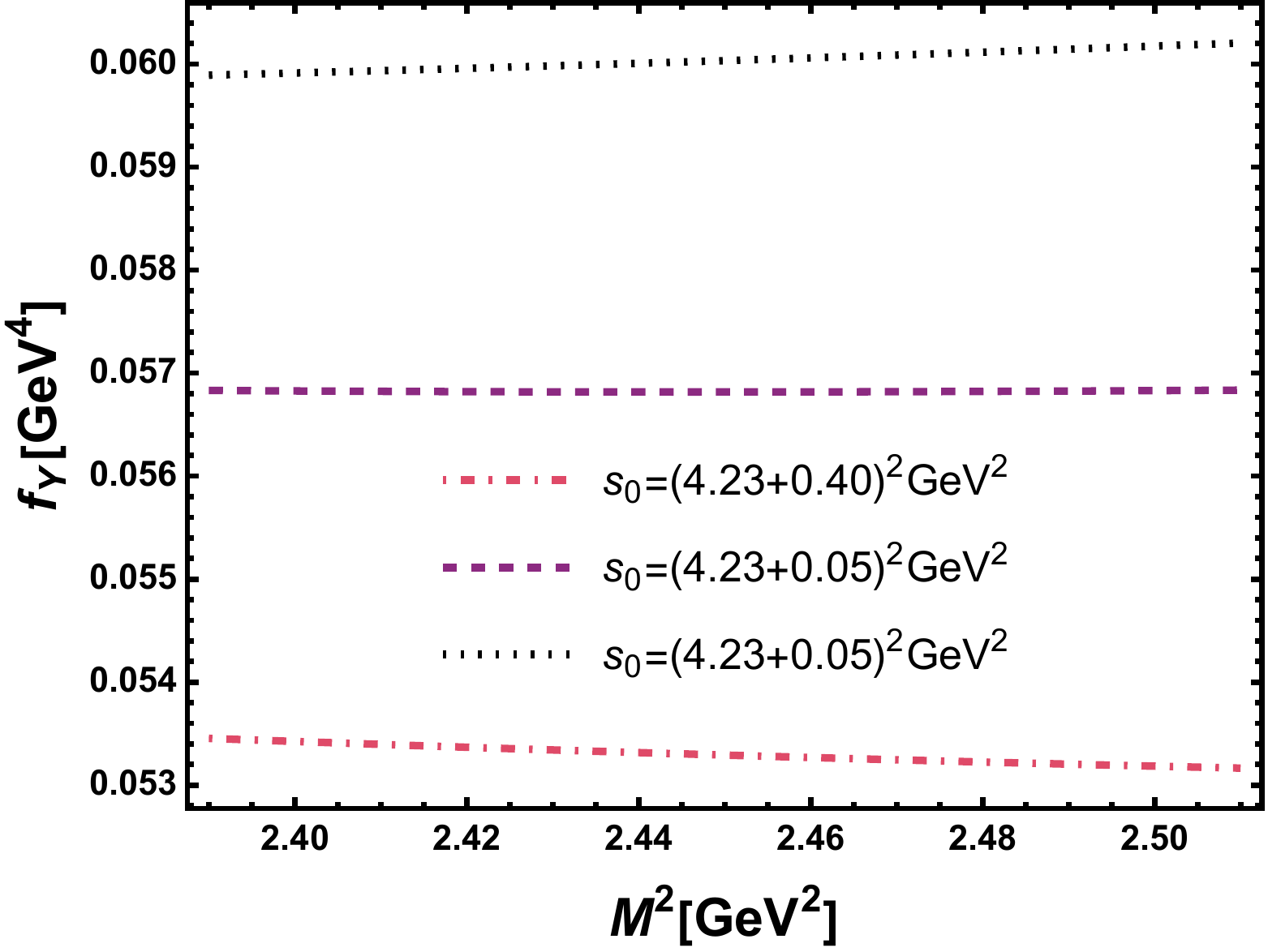}
  \caption{The mass [first] and the decay constant of $Y(4230)$ [second] as a function of
the Borel parameter $M^2$ at different fixed values of $s_0$.}
  \label{Fig:fig2}
\end{figure}

The outcomes of the mass $m_Y$ and the decay constant $f_Y$ as functions of the parameters $M2$ are shown in FIG.\ref{Fig:fig2}.
The orange shape in the first picture of FIG.\ref{Fig:fig2} corresponds to the measurements taken by the Belle Collaboration\cite{BESIII:2016bnd}.
The other curves show our prediction at fixed $s_0\in\{(4.23+0.40)^2,(4.23+0.50)^2,(4.23+0.60)^2\}$. our prediction can be consistent with the measurement.
At a fixed point of $M^2=2.45$ $\rm GeV^2$, our result for the mass reads
\begin{equation}
	\begin{aligned}
		\quad m_Y=4.22^{+0.08}_{-0.07}\ \text{GeV}.
	\end{aligned}
\end{equation}
Our mass prediction shows that the generalization of our method is valid.
We then extend the method to evaluate the decay constant of $Y(4230)$, the result at the same typical point reads
\begin{equation}
	\begin{aligned}
		f_Y=(0.0568\pm0.003)\ \text{GeV}^4.
	\end{aligned}
\end{equation}
The mass and decay constant be input parameters to calculate the decay width of $\psi$(4230)$\to J/\psi f_0(980)$.

The $\psi$(4230) branching ratios from PDG\cite{ParticleDataGroup:2020ssz}shows that
\begin{equation}
	\begin{aligned}
		\frac{\Gamma(J/\psi f_0(980),f_0(980)\to \pi^+\pi^-)}{\Gamma(J/\psi\pi^+\pi^-)}=0.17\pm0.13.
	\end{aligned}
\end{equation}
We can estimate the upper limit of $\Gamma(J/\psi f_0(980)$, $f_0(980)\to \pi^+\pi^-)$,
by assuming that $\psi\to$ $J/\psi\pi^+\pi^-$ is the only decay process of $\psi(4230)$
With the width of $\Gamma_{\psi}=76.6\pm14.2\pm2.4$ MeV,
we can obtain
\begin{equation}
	\begin{aligned}
		&\Gamma(\psi\to J/\psi f_0(980),f_0(980)\to \pi^+\pi^-)=\\
		&\Gamma(\psi\to J/\psi f_0(980))\mathcal{B}(f_0(980)\to \pi^+\pi^-)<13.022\ \text{MeV}.
	\end{aligned}
\end{equation}

Also from PDG one can find from the $f_0$(980) branching ratios that
\begin{equation}
\begin{aligned}
	\frac{\Gamma(\pi\pi)}{(\Gamma(\pi\pi)+\Gamma(K\bar{K}))}=0.75^{+0.11}_{-0.13},
\end{aligned}
\end{equation}
and the partial width that
\begin{equation}\label{gamma}
\begin{aligned}
	P(f_0(980)\to \gamma\gamma)\equiv\frac{\Gamma(f_0(980)\to \gamma\gamma)}{\Gamma(f_0(980))}=0.31^{+0.05}_{-0.04}.
\end{aligned}
\end{equation}
The process $f_0(980)\to \pi\pi$, $f_0(980)\to K\bar{K}$ and $f_0(980)\to \gamma\gamma$
are the main decay process of $f_0$(980). From Eq.\eqref{gamma} comes out the
partial width $P(f_0(980)\to\Gamma(\pi\pi)+\Gamma(K\bar{K}))=0.69^{+0.04}_{-0.05}$.
So we estimate
\begin{equation}
	\begin{aligned}
		&\mathcal{B}(f_0(980)\to \pi^+\pi^-)\\ &\approx0.75^{+0.11}_{-0.13}
		\times P(f_0(980)\to\Gamma(\pi\pi)+\Gamma(K\bar{K}))\approx0.52.
	\end{aligned}
\end{equation}

Then we can finally conclude that
\begin{equation}
	\begin{aligned}
		\Gamma(\psi \to J/\psi f_0(980))<25.0\ \text{MeV}.
	\end{aligned}
\end{equation}

\begin{figure}[htbp]
  \includegraphics[width=7cm]{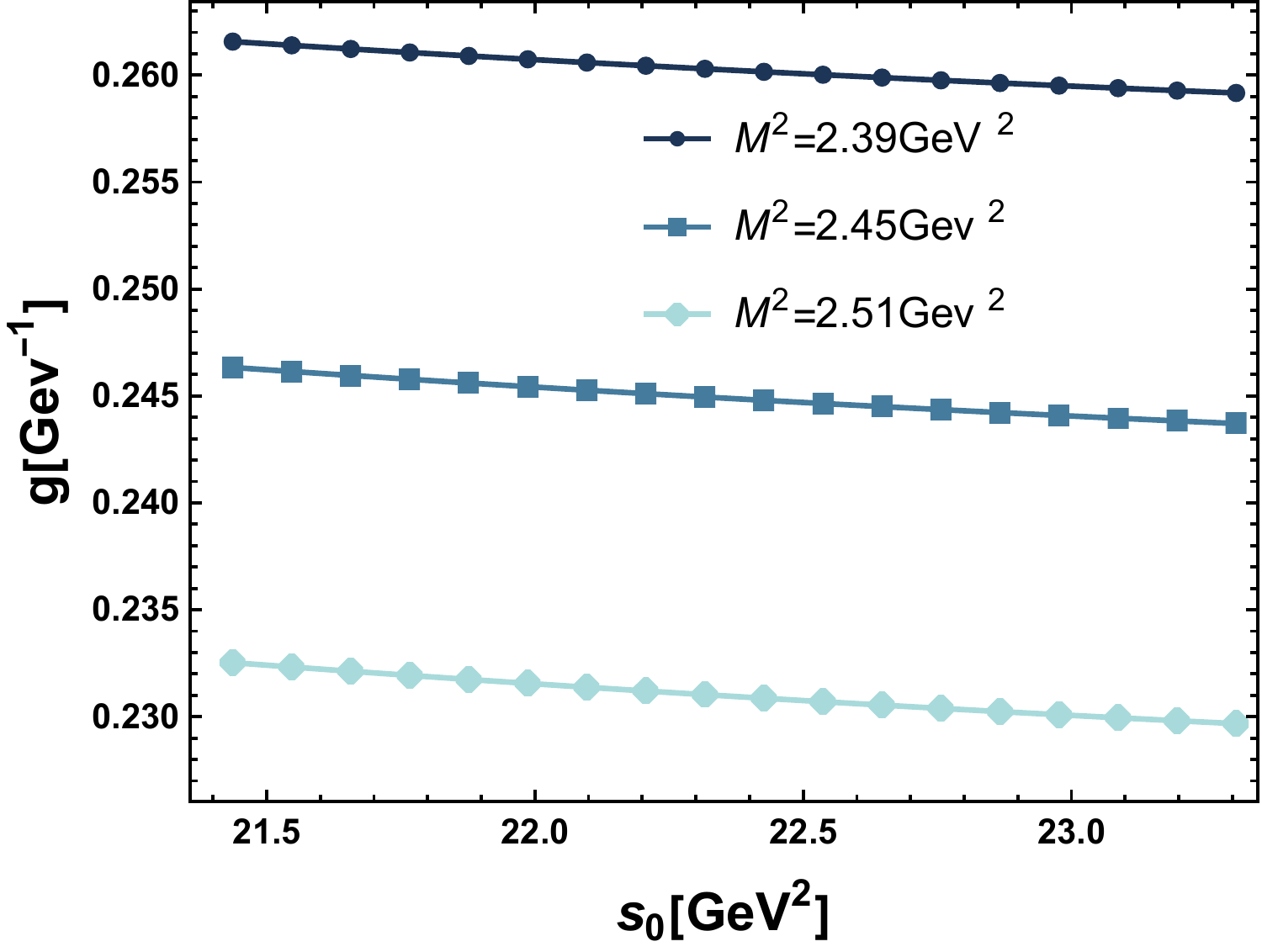}
  \caption{The strong coupling $g_{YJ/\psi f_0(980)}$ as a function of the threshold parameter $s_0$ at different fixed values of $M^2$}
  \label{Fig:f0}
\end{figure}
\begin{figure}[htbp]
  \includegraphics[width=7cm]{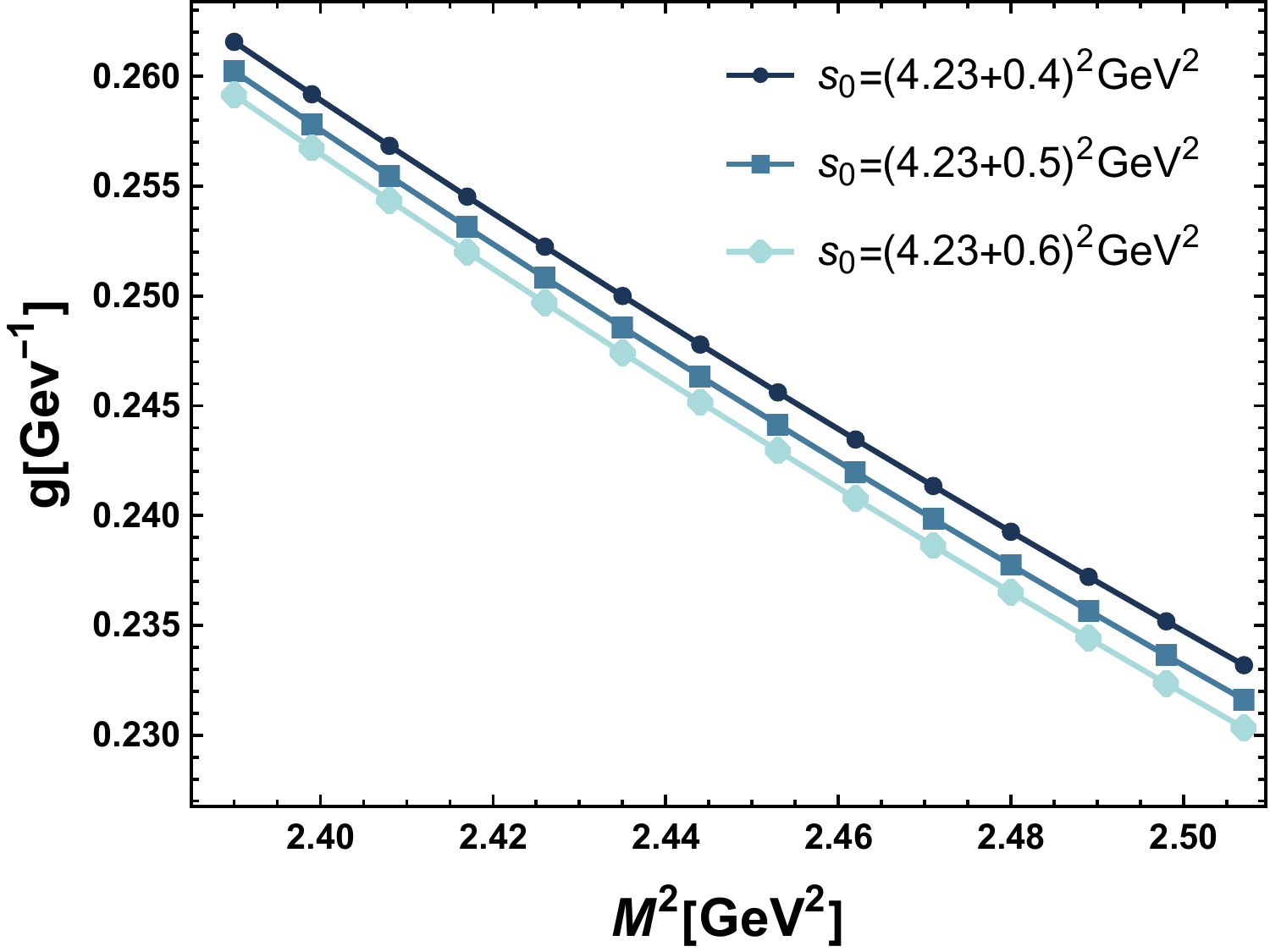}
  \caption{The strong coupling $g_{YJ/\psi f_0(980)}$ as a function of the Borel parameter $M^2$ at different fixed values of $s_0$}
  \label{Fig:f01}
\end{figure}
As we seen in Fig.\ref{Fig:f01}, the blue,
green and black curves show clear dependence of our prediction on $s_0$ and $M^2$. 
For $M^2$ and $s_0$, we use the same values as in the analysis of the mass. 
The result is shown in Fig.\ref{Fig:f0} and Fig.\ref{Fig:f01}.
By choosing appropriate parameters, our prediction for $g_{YJ/\psi f_0}$ is
\begin{equation}
	\begin{aligned}
		g_{YJ/\psi f_0}=(0.245\pm 0.01)\ \text{GeV}^{-1}.
	\end{aligned}
\end{equation}
Where we take the average result of $g_{YJ/\psi f_0}$.
The width of this decay can be obtained by
means of Eq.\eqref{couplingn}
\begin{equation}
	\begin{aligned}
		\Gamma(Y \rightarrow J/\psi f_0(980))=(1.28\pm 0.1)\ \text{MeV},
	\end{aligned}
\end{equation}
which is less than the upper limit of $\psi$(4230) $\to J/\psi f_0(980)$ decay width.
Combining this result with the prediction result of mass, we may conclude that $Y$(4230) could be a tetraquark state.
But, since the lack of experiments data about $\psi$(4230) $\to J/\psi f_0(980)$ decay width,
we still need future experiments to further determine whether $Y$(4230) have the possibility be a tetraquark state or not.

\section{Summary}
\label{sec:summary}

In this research, we designate $Y$(4230) as a vector tetraquark state in order to concurrently analyze $Y(4230)$'s mass,
decay constant, and decay into $J/\psi f_0(980)$. 
The mass of $Y(4230)$ is evaluated through a different calculation technique developed in two-point sum rules,
and the result is in agreement with the mass of $Y(4230)$ in PDG. 
Then we extend the technique to calculate the decay constant of $Y(4230)$.
Using the method of light cone sum rules, we calculate the coupling constant $g_{YJ/\psi f_0}$
and discover the result for $Y\to J/\psi f_0(980)$ decay width.
Then we assume that $J/\psi\pi\pi $ is the most significant channel, overwhelming all the other channels. 
Therefore we can consider the width of $\psi(4230)$ as the width of $J/\psi \pi\pi $. 
Since we know the branching ratios of $\Gamma(J/\psi f_0(980),f_0(980)\to \pi^+\pi^-)/\Gamma(J/\psi\pi^+\pi^-) $from PDG,
we can estimate the upper limit of $J/\psi f_0$ channel.
The decay width of $Y\to J/\psi f_0(980)$ is less than the upper limit. 
Our prediction of the mass of $Y(4230)$ is in agreement with that of $Y(4230)$ in PDG and the decay width of $Y(4230)\to J/\psi f_0(980)$ does not exceed its theoretical limits. There is a possibility that $Y$(4230) could be a tetraquark.
In the future, experiments will be more helpful in determining whether or not this structure of $Y(4230)$ is appropriate.

\begin{acknowledgments}
Hao Sun is supported by the National Natural Science Foundation of China (Grant No.12075043).
\end{acknowledgments}

\section{Appendix}

\subsection{Particle distribution amplitudes}
\label{appendix:A}

The matrix elements of the $f_0$ can be expanded in terms of the corresponding distribution amplitudes.
Below we provide expressions for $\braket{f_0(980)(q)|\bar{d}(x)\Gamma^au(0)|0}$\cite{Colangelo:2010bg}:
\begin{equation}
	\begin{aligned}
		&\braket{f_0(q)|\bar{s}(x)\gamma_\mu s(0)|0}=\bar{f}_{f_0}q_\mu\int_0^1due^{iuq\cdot x}\Phi_{f_0}(u),\\
		&\braket{f_0(q)|\bar{s}(x)s(0)|0}=m_{f_0}\bar{f}_{f_0}\int_0^1due^{iuq\cdot x}\Phi_{f_0}^s(u),\\
		&\braket{f_0(q)|\bar{s}(x)\sigma_{\mu\nu} s(0)|0}\\
		&=-\frac{m_{f_0}}{6}\bar{f}_{f_0}(q_\mu x_\nu-q_\nu x_\mu)\int_0^1due^{iuq\cdot x}\Phi_{f_0}^\sigma(u),
	\end{aligned}
\end{equation}
where the LCDA $\Phi_{f_0}$ is twist-2 light-cone distribution amplitudes of $f_0(980)$, and the other two are twist-3 distribution amplitudes.
Meantime, we use the following normalization
\begin{equation}
	\begin{aligned}
		&\int_0^1du\Phi_{f_0}(u)=0,\int_0^1du\Phi_{f_0}^s(u)=1,\int_0^1du\Phi_{f_0}^\sigma(u)=1.
	\end{aligned}
\end{equation}

\subsection{The formula for LCSR}
\label{appendix:B}

When calculate the OPE part of the correlation function, 
we will encounter various four-dimensional integrals in the momentum spaces. 
Before performing the integration, it is often to use the Feynman's parametric integral formula:
\begin{equation}
	\begin{aligned}
		&\frac{1}{A_1^{a_1}\cdot\cdot\cdot A^{a_n}_n}=\frac{\Gamma(a_1\cdot\cdot\cdot a_n)}{\Gamma(a_1)\cdot\cdot\cdot\Gamma(a_n)}\int_0^1dx_1\cdot\cdot\cdot\int_0^1dx_n\\
		&\frac{\delta(1-x_1+\cdot\cdot\cdot+x_n)x_1^{a_1-1}\cdot x_2^{a_2-1}\cdot\cdot\cdot x_n^{a_n-1}}{(x_1A_1+\cdot\cdot\cdot+x_nA_n)^{a_1+\cdot\cdot\cdot a_n}}.
	\end{aligned}
\end{equation}
In general, Feynman integrals contain:
\begin{equation}\label{fe}
	\begin{aligned}
		I(D;a,q)=\int\frac{d^Dp}{(2\pi)^D}\frac{1}{(p^2+2pq+m^2)^a}.
	\end{aligned}
\end{equation}
This integral can be reduced to
\begin{equation}
	\begin{aligned}
		I(D;a,q)&=\int\frac{d^Dp}{(2\pi)^D}\frac{1}{(p^2+2pq+m^2)^a}\\
		&=\frac{1}{(4\pi)^{D/2}}\frac{\Gamma(a-D/2)}{\Gamma(a)}(m^2-q^2)^{D/2-a}.
	\end{aligned}
\end{equation}
To obtain a formula in proportion to $p_\mu$ like 
\begin{equation}
	\begin{aligned}
		I(D;a,q)&=\int\frac{d^Dp}{(2\pi)^D}\frac{p_\mu}{(p^2+2pq+m^2)^a}\\
		&=\frac{1}{(4\pi)^{D/2}}\frac{\Gamma(a-D/2)}{\Gamma(a)}\frac{q_\mu}{(m^2-q^2)^{-D/2+a}},
	\end{aligned}
\end{equation}
we can differentiate equation Eq.\eqref{fe} with momentum $q$ one time. The higher tensors $p_\mu\cdots p_\nu$ in the
integrand come form higher differentiations.
Now above equation encounter a pole in Gamma function when dimension $D\to 4$, i.e. $\Gamma(0)\to\infty$.
We can use equation
\begin{equation}
	\begin{aligned}
		\frac{\Gamma(n-\frac{1}{2}d)}{a^{2n-d}}=\int_0^\infty d\lambda\lambda^{n-d/2-1}e^{-\lambda a^2}
	\end{aligned}
\end{equation}
to get rid of Gamma function and perform the replacement
\begin{equation}
	\begin{aligned}
		\int_0^\infty \frac{d\lambda}{\lambda^n}e^{-\lambda K}\to(-1)^n\frac{K^{n-1}lnK}{(n-1)!}.
	\end{aligned}
\end{equation}
To obtain the final expression of the correlation function, 
we need the imaginary part of results and the integration over the Feynman parameters.

\subsection{The formula for mass and decay constants}
\label{C}

Here we present calculation details of integral when we dealing with Eq.\eqref{propagator}.
we need to consider a general integral
\begin{equation}\label{A}
	\begin{aligned}
		&I(p^2)=\int d^4x  e^{ipx} \frac{1}{(x^2)^n} \frac{K_i(m_1\sqrt{-x^2})}{(\sqrt{-x^2})^i}\frac{K_j(m_2\sqrt{-x^2})}{(\sqrt{-x^2})^j}.
	\end{aligned}
\end{equation}
Use the integral representation of Bessel function
\begin{equation}
	\begin{aligned}
		\frac{K_i(m_1\sqrt{-x^2})}{(\sqrt{-x^2})^i}=\int_0^\infty\frac{dt}{t^{i+1}}\exp{[-\frac{m_1}{2}(t-\frac{x^2}{t})]},
	\end{aligned}
\end{equation}
we have
\begin{equation}
\begin{aligned}
I=&\frac{1}{4}\int d^4x e^{ipx}\frac{1}{(x^2)^n}\int_0^\infty\frac{dt_1}{t_1^{i+1}}\exp{[-\frac{m_1}{2}(t_1-\frac{x^2}{t_1} ) ]}\\
&\times\int_0^\infty\frac{dt_2}{t_2^{j+1}}\exp{[-\frac{m_2}{2}(t_2-\frac{x^2}{t_2} ) ]}\\
=&\frac{i(-1)^n}{4\Gamma(n)}\int d^4x e^{ipx}\int_0^\infty d\lambda \lambda^{n-1} \exp(-\lambda x^2 )\int_0^\infty\frac{dt_1}{t_1^{i+1}}\\
 &\times\exp{[-\frac{m_1}{2}(t_1+\frac{x^2}{t_1} ) ]} \int_0^\infty\frac{dt_2}{t_2^{j+1}}\exp{[-\frac{m_2}{2}(t_2+\frac{x^2}{t_2} ) ]}.
\end{aligned}
\end{equation}

Introduce new variables
\begin{equation}
a=\frac{2m_1}{t_1},\ b=\frac{2m_2}{t_2} \to dt_1 =-\frac{2m_1}{a^2}da,\ dt_2 =-\frac{2m_2}{b^2}db,
\end{equation}
leading to equation
\begin{equation}
\begin{aligned}
I&=\frac{(-1)^ni}{4 \Gamma(n)}\frac{1}{(2m_1)^{i}(2m_2)^j} \int_0^\infty  da\ a^{i-1} \int_0^\infty  db\ b^{j-1}  \int_0^\infty d\lambda \lambda^{n-1}\\
	&\times\exp[ -(\frac{m_1^2}{a}+\frac{m_2^2}{b} ) ] \int d^4x \ \exp{[-\frac{1}{4}(a+b+4\lambda)x^2-ipx ]}\\
&=\frac{(-1)^ni}{4 \Gamma(n)}\frac{16\pi^2}{(2m_1)^{i}(2m_2)^j} \int_0^\infty  da \int_0^\infty  db\ a^{i-1} b^{j-1}  \int_0^\infty d\lambda \lambda^{n-1}\\
&\times\frac{1}{(a+b+4\lambda)^2} \exp[ \frac{-p^2}{(a+b+4\lambda)} ] \exp[ -(\frac{m_1^2}{a}+\frac{m_2^2}{b} ) ].
\end{aligned}
\end{equation}
Than perform $a\to4a$, $b\to4b$, we obtain
\begin{equation}
\begin{aligned}
I=&\frac{(-1)^ni}{\Gamma(n)}\frac{2^{i+j-2}\pi^2}{(m_1)^{i}(m_2)^j} \int_0^\infty  da \int_0^\infty  db\ a^{i-1} b^{j-1}  \int_0^\infty d\lambda \lambda^{n-1}\\
&\times  \frac{1}{(a+b+\lambda)^2} \exp[ \frac{-p^2}{4(a+b+\lambda)} ] \exp[ -(\frac{m_1^2}{4a}+\frac{m_2^2}{4b} ) ].
\end{aligned}
\end{equation}
Now we introduce the variables $\rho$, $x$, and $y$, defined by
\begin{equation}
	\begin{aligned}
		\rho=(a+b+\lambda), x=\frac{a}{a+b+\lambda},y=\frac{b}{a+b+\lambda}.
	\end{aligned}
\end{equation}
Then we have
\begin{equation}
	\begin{aligned}
	\mbox{d}\lambda\mbox{d}a\mbox{d}b&=
		\begin{vmatrix}
		\frac{\partial\lambda}{\partial\rho} & \frac{\partial a}{\partial x} &\frac{\partial b}{\partial y} \\
		\frac{\partial\lambda}{\partial\rho} & \frac{\partial a}{\partial x} &\frac{\partial b}{\partial y}\\
		\frac{\partial\lambda}{\partial\rho} & \frac{\partial a}{\partial x} &\frac{\partial b}{\partial y}
\end{vmatrix}
\mbox{d}\rho\mbox{d}x\mbox{d}y\\
&=\rho^2 \mbox{d}\rho\mbox{d}x\mbox{d}y,
	\end{aligned}
\end{equation}
which leads to
\begin{equation}
	\begin{aligned}
			&I=\frac{(-1)^ni}{\Gamma(n)}\frac{2^{i+j-2}\pi^2}{(m_1)^{i}(m_2)^j} \int du\varphi(u)\int_0^\infty d\rho\int_0^1dx\int_0^1dy(\rho x)^{i-1} \\
			&\times (\rho y)^{j-1}(\rho(1-x-y))^{n-1}\exp[ \frac{-p^2}{4\rho} ] \exp[ -(\frac{m_1^2}{4\rho x}+\frac{m_2^2}{4\rho y} ) ].
	\end{aligned}
\end{equation}
Applying the double Borel transformations with respect to $-p^2\to M^2$, we obtain
\begin{equation}
	\begin{aligned}
		&\tilde{I}(M^2)=\frac{(-1)^ni}{\Gamma(n)}\frac{2^{i+j-2}\pi^2}{(m_1)^{i}(m_2)^j} \int_0^\infty d\rho\int_0^1  dx \int_0^1  dy (\rho x)^{i-1}\\
&\times (\rho y)^{j-1} (\rho(1-x-y))^{n-1}\delta(\frac{1}{M^2}-\frac{1}{4\rho})\exp[ -(\frac{m_1^2}{4\rho x}+\frac{m_2^2}{4\rho y} ) ]\\
&=\frac{(-1)^{n}i}{\Gamma(n)}\frac{2^{2-2n-i-j}\pi^2}{(m_1)^{i}(m_2)^j}(M^2)^{i+j+n-1}\int_0^1  dx \int_0^1  dy \\
&\times x^{i-1} y^{j-1} (1-x-y)^{n-1}\exp[ -(\frac{m_1^2}{M^2x}+\frac{m_2^2}{M^2 y} ) ],
	\end{aligned}
\end{equation}
where $\rho_0=\frac{M^2}{4}$.

Introducing new variables, $\sigma_i=\frac{1}{M_i^2}$, we have
\begin{equation}
	\begin{aligned}
			\tilde{I}(\sigma)
&=\frac{(-1)^{n}i}{\Gamma(n)}\frac{2^{2-2n-i-j}\pi^2}{(m_1)^{i}(m_2)^j}\frac{1}{\sigma^{i+j+n-1}}\int_0^1  dx \int_0^1  dy \\
&\times x^{i-1} y^{j-1} (1-x-y)^{n-1} \exp[ -(\frac{m_1^2}{x}+\frac{m_2^2}{y} )\sigma ]\\
&=\frac{C}{\Gamma(i+j+n-1)}\int_0^1  dx \int_0^1  dy x^{i-1} y^{j-1} (1-x-y)^{n-1}\\
&\times \exp[ -(\frac{m_1^2}{x}+\frac{m_2^2}{y} )\sigma]\int_0^\infty \mbox{d}z\exp[-z\sigma]z^{i+j+n-2}\\
&=\frac{C}{\Gamma(i+j+n-1)}\int_0^1  dx \int_0^1  dy x^{i-1} y^{j-1} (1-x-y)^{n-1}\\
&\times \int_0^\infty \mbox{d}z\exp[-(z+\frac{m_1^2}{x}+\frac{m_2^2}{y} )\sigma]z^{i+j+n-2}.
	\end{aligned}
\end{equation}
Where $C=\frac{(-1)^{n}i}{\Gamma(n)}\frac{2^{2-2n-i-j}\pi^2}{(m_1)^{i}(m_2)^j}$.
Applying double Borel transformation with respect to $\sigma\to \frac{1}{s}$,
we obtain the spectral density
\begin{equation}
	\begin{aligned}
			\rho(s)
&=\frac{C}{\Gamma(i+j+n-1)}\int_0^1  dx \int_0^1  dy x^{i-1} y^{j-1} (1-x-y)^{n-1}\\
&\times \int_0^\infty \mbox{d}z\delta[s-(z+\frac{m_1^2}{x}+\frac{m_2^2}{y} )]z^{i+j+n-2}\\
&=\frac{C}{\Gamma(i+j+n-1)}\int_0^1  dx \int_0^1  dy x^{i-1} y^{j-1} (1-x-y)^{n-1}\\
&\times (s-\frac{m_1^2}{x}-\frac{m_2^2}{y})^{i+j+n-2}\theta(s-\frac{m_1^2}{x}-\frac{m_2^2}{y}).
	\end{aligned}
\end{equation}

Similarly, we also need to consider integral
\begin{equation}
\begin{aligned}
I&=\int d^4x  e^{ipx} \frac{x^\mu x^\nu}{(x^2)^n} \frac{K_i(m_1\sqrt{-x^2})}{(\sqrt{-x^2})^i}\frac{K_j(m_2\sqrt{-x^2})}{(\sqrt{-x^2})^j}\\
&=-\partial_\mu^{(p)}\partial_\nu^{(p)}\int d^4x  e^{ipx} \frac{1}{(x^2)^n} \frac{K_i(m_1\sqrt{-x^2})}{(\sqrt{-x^2})^i}\frac{K_j(m_2\sqrt{-x^2})}{(\sqrt{-x^2})^j},
\end{aligned}
\end{equation}
and derived spectral density
\begin{equation}
	\begin{aligned}
		\rho(s)
&=\frac{C_1}{\Gamma(i+j+n+1)}\int_0^1  dx \int_0^1  dy x^{i-1} y^{j-1} (1-x-y)^{n-1}\\
&\times (s-\frac{m_1^2}{x}-\frac{m_2^2}{y})^{i+j+n}\theta(s-\frac{m_1^2}{x}-\frac{m_2^2}{y})p_\mu p_\nu\\
&+\frac{C_2}{\Gamma(i+j+n)}\int_0^1  dx \int_0^1  dy x^{i-1} y^{j-1} (1-x-y)^{n-1}\\
&\times (s-\frac{m_1^2}{x}+\frac{m_2^2}{y})^{i+j+n-1}\theta(s-\frac{m_1^2}{x}-\frac{m_2^2}{y})g_{\mu\nu},
	\end{aligned}
\end{equation}
Where $C_1=\frac{(-1)^{n+1}i}{\Gamma(n)}\frac{2^{4-2n-i-j}\pi^2}{(m_1)^{i}(m_2)^j}$, $C_2=\frac{(-1)^{n+2}i}{\Gamma(n)}\frac{2^{3-2n-i-j}\pi^2}{(m_1)^{i}(m_2)^j}$.

\bibliography{ref}

\end{document}